# Investigating Assumptions and Proposals for Blockchain Integration in the Circular Economy. A Delphi Study.

Giulio Caldarelli

Department of Management

University of Turin

giulio.caldarelli@unito.it

v0.2

## Abstract

Given the rising interest in the circular economy and blockchain hype, numerous integrations were proposed. However, studies on the practical feasibility were scarce, and the assumptions of blockchain potential in the circular economy were rarely questioned. With the help of eleven of the most prominent blockchain experts, the present study critically analyzed technology integration in many areas of the circular economy to forecast their possible outcomes. Delphi's technique is leveraged to reach a consensus among experts' visions and opinions. Results support the view that some circular economy integrations are unlikely to succeed, while others if specific conditions are met, may prove to be successful in the long run.

Keywords: *Blockchain, Smart Contracts, Oracles, Circular Economy, Sustainability, Delphi study*.

### 1. Introduction

"a systematic assessment of advantages and disadvantages of blockchain for a circular economy is lacking for different use cases and contexts...we argue that critical dialogue led by industry is essential" [1]. This interesting paper by Böckel et al. sheds important light on the fact that research on blockchain potential in the circular economy lacks a more technical and critical analysis led by experts from the industry. The researchers argue that only general statements regarding the potential of blockchain for a circular and sustainable economy are currently made without further systematic approaches. Literature on blockchain and circular economy advocates technology integration intending to improve traceability and trust among actors, fostering adoption and enhancement of circular practices [2], [3]. However, further studies on technical aspects that show how the technology is supposed to achieve those claims are lacking. As much research shows, the load of hype about blockchain is fueled with false expectations and misconceptions, of which the most common concerns the ability to monitor external states or to provide a unique source of truth [4]–[6]. Unclarity and misconceptions may easily create biases when theory and research are built with regard to real-world integrations. Therefore, as advocated by Böckel et al. a transparent and unbiased background is necessary to build further theoretical and practical technology integration in the real world.

The scope of the present study is to investigate blockchain assumptions for circular economy with the contribution of prominent experts in the field of blockchain technology. In order to reach an agreement among their views, the Delphi methodology is leveraged and adapted for this specific research. A literature review is performed to obtain the items for the Delphi study. Experts then evaluated items during direct interviews, where further comments were collected to understand their opinions on each topic better. A final survey is then conducted to reach a final consensus.

Results from this research provide insights into the potential of blockchain in enhancing transparency and trust in some CE practices. Thanks to experts' technical comments and examples, a thorough explanation is also provided for blockchain's potential to incentivize green practice. Unsupported blockchain integration assumptions are also thoroughly explained to better understand if they are based on misconceptions, technology limitations, or conceptual weaknesses. Thanks to further suggestions by experts, a detailed list of CE applications that are expected to benefit from blockchain integration is also provided.

The objectives of the paper are the following:

1) Provide a background of blockchain technology that can be leveraged to build research on real-world integrations such as circular economy as requested by Böckel et al. [1]
2) Find common assumptions and proposals of blockchain and circular economy on which literature is based.
3) Investigate specific limitations and delimitations of each assumption.
4) Produce a list of applications for which blockchain is expected to contribute to CE practices, enlightening those unlikely to support.

The paper proceeds as follows. Section two provides a background of Blockchain technology that should help understand strengths and limitations in real-world applications as well as a background of circular economy to better contextualize the study. Section three introduces the methodology, explaining the adaptations for this study and the expert composition. Section four presents the results and discusses the quantitative part, while section five provides an overview and discussion of the qualitative part, outlining concerns and recommendations from experts for each topic. Section 6 summarizes and concludes the paper by providing hints for further research.

## 2. Theoretical Background

### 2.1 Circular economy

The concept of circular economy has emerged as a sustainable development approach sought to overcome the challenges posed by the traditional take-make-dispose system. Recognizing the finite nature of resources, circular economy seeks to establish a regenerative and restorative economic system.

Over the years, the circular economy concept has gained significant traction and has become a pivotal topic in the academic and practitioner fields. The backbone of this concept is the principle of reducing material consumption and waste generation while maximizing the utility of resources through strategies such as recycling, refurbishing, and remanufacturing [7]. The circular economic model tries to decouple economic growth from resource consumption and environment degradation [8].

The benefits of a circular economic model have been investigated and acknowledged across environmental, economic, and social dimensions. From an environmental point of view, [9] states that a circular economy has the potential to alleviate pressure on natural ecosystems, reduce greenhouse gas emissions, and conserve finite resources. In an economic context, instead, integrating circular practices stimulates innovation, enhances resource efficiency, and fosters job creation [10]. Finally, from a societal perspective, the circular economy model should encourage customers to adopt sustainable consumption practices, encouraging behavior that values durability, repairability, and product longevity over disposability [11], [12].

While the circular economy's potential is vast, its practical implementation faces various challenges. Most of all, the lack of information throughout the products' lifecycle constitutes a significant barrier to the effective implementation of circular economy principles [13]. Many believe that integrating blockchain technology may help reduce the critical lack of data and facilitate the overall transition to a circular economic model [14]–[18].

## 2.2 Blockchain technology and its characteristics

Blockchain is a distributed ledger whose data is added through a consensus mechanism and contained in blocks. Blockchain characteristics and purposes are considerably heterogeneous. Some divide blockchains into public and private based on the access type, with some hybrid forms recognized as consortium [19], [20]. Other works distinguish blockchains by ecosystems such as Bitcoin, Ethereum, or Hyperledger, enumerating their main characteristics [21], [22]. Arguably, both a distinction for access type and by ecosystem cannot provide the necessary background to understand and better link real-world applications (such as those for circular economy) with blockchain technology. However, a distorted view of integration potential is generated if blockchain characteristics are not sufficiently and efficiently defined. For example, if research proposes to improve sustainable supply chains with blockchain considering Bitcoin characteristics, a false expectation is generated if research or practical application is made by leveraging alternative chains.

Given the scope of the study and to clarify real-world blockchain integration for CE practices, an approach based on the blockchain trilemma [23] is perceived as appropriate as it efficiently describes blockchain's strengths and weaknesses.

### 2.2.1 Blockchain trilemma

A distributed ledger such as Bitcoin is renowned for its high level of Decentralization and Security. Bitcoin, however, is also known for being not scalable. It means that the number of transactions that can be processed is low compared to financial circuits such as Visa and Mastercard [24]. Basically, in order to reach a high level of decentralization and security, Bitcoin had to renounce to a considerable level of throughput. Alternative blockchains, such as Ethereum or Tron, could offer a high level of scalability by giving up a consistent level of decentralization and security. It is widely recognized that maximizing decentralization security and scalability within the same blockchain is not technically feasible. Maximizing one of these characteristics requires renouncing one or both of the other two [23], [25]–[27]. The existence of the trilemma can create a paradox when blockchain has to be integrated with real-world applications. Real-world applications may require, for example, a high level of scalability for a blockchain to be technically implemented. However, that implies that the security and decentralization of the resulting integration will be limited. If the initial scope for integrating blockchain were to increase security and decentralization, then the outcome would probably not satisfy the expectations. Therefore, given blockchain's heterogeneity, the integration into legacy business cannot imply any increase in decentralization, security, or scalability. A graphical overview of the trilemma is shown in Figure 1.

Figure 1. Blockchain Trilemma

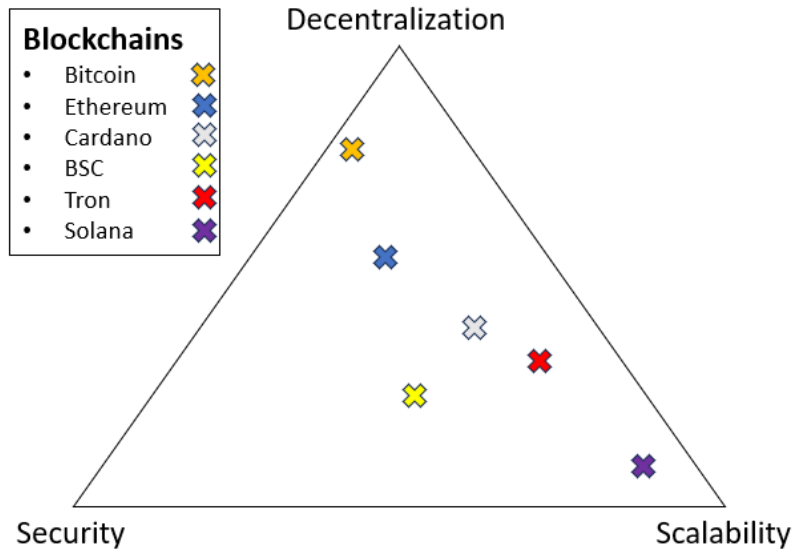

*Author Elaboration

### 2.3 Consensus and real-world Data on the Blockchain

A strict form of consensus ensures the decentralization and security of Bitcoin. In fact, in order to add a block to the Bitcoin blockchain, a cryptographic puzzle has to be solved, which requires a considerable amount of investment in hardware and energy [28]. If other users can verify the genuineness of the procedure, the agent that has added a new block is rewarded with newly issued bitcoins. The security of the ledger is then guaranteed by the fact that if someone produces a block that will not be recognized as valid, it will not be rewarded, and the mining cost incurred is forfeited. This consensus type is known as proof-of-work and consumes wealth whether the block produced is considered valid or not. An alternative consensus type is based on the vote of some agents willing to put their wealth as a guarantee that the block they are producing is valid. If the block is recognized as valid, the agents are rewarded; otherwise, the wealth put at stake is slashed. This consensus type is known as proof-of-stake, and those producing blocks consume wealth only in the case their block is considered invalid as a form of punishment [29].

It Is then Imperative to understand that not all the data on the blockchain Is verified by the consensus mechanism but only the one that is strictly required to add a new block. To make an example, on Bitcoin the data verified by the consensus that can then be considered secure and reliable is the one concerning the transfer of the bitcoin cryptocurrency [28]. If the Bitcoin blockchain is used as a ledger to register or store academic transcripts, credentials, residency permits, or traceability data, for example, the Bitcoin consensus mechanism does not verify the authenticity of this data since is unrelated to bitcoin transactions [30], [31]. Although leveraging different consensus mechanisms, the same can be said for other blockchain types. Therefore, Bitcoin or other blockchains generally have no means to guarantee the authenticity of the data stored.

To address this issue, researchers and practitioners are trying to develop systems known as "oracles" to ensure the reliability of real-world data transferred to the blockchain, leveraging consensus mechanisms that tend to replicate those of blockchains. Given the issue's complexity, proposed solutions are highly controversial, and a robust oracle is still unavailable [32], [33].

Some oracle solutions ensure that the data registered to the blockchain is not manipulated in the transfer process but do not guarantee its authenticity at the source [34]. Other oracle solutions try to leverage the wisdom of the crowd to provide blockchain with data that is widely recognized as accurate [35]. However, it only works for information that is widely accessible, such as weather, exchange rates, election results, game scores, and so on. Mechanisms for ensuring the truthfulness of data that is not widely available or disclosed, such as personal, product, or company information, are still under debate. Arguably blockchain alone can efficiently store data but is unable to gather data, and therefore, it can hardly be used to check or certify data for applications that require actions or intervention in the real-world, having no direct access to that [36], [37].

Merely integrating blockchain in real-world applications such as traceability of waste or carbon emission, therefore, does not give any ulterior guarantee on the authenticity and reliability of data provided unless a clear method of data acquisition and verification that should also leverage other technologies (i.e., IoT) is exhaustively provided and explained [38]. Table 1 summarizes information provided in this section.

Table 1. Concepts to be aware of when evaluating blockchain real-world integrations*.

| Concept | Description | Reference |
|---|---|---|
| Trilemma | Blockchains tend to balance Scalability, Security, and Decentralization. If one characteristic is maxed, the others are limited. | [23], [25], [26] |
| Consensus | The consensus mechanism only verifies the information strictly necessary to produce a new block. | [28], [29] |
| Real-world data on a blockchain | Virtually any real-world data can be stored on a blockchain as with ordinary databases. No quality criteria are required for extrinsic data to be uploaded on a blockchain. | [30], [31], [39] |
| Secure Oracles | Highly secure oracles ensure that the data transmitted is not manipulated. They can also provide salient information. To date, they are still far from being truth machines. | [34], [35], [40], [41] |

*Author elaboration

## 3. Methodology

Given the scope of the study, the author opted for a Delphi. The Delphi method [42] is widely utilized for achieving convergence of opinions concerning real-world knowledge solicited from experts in certain topics [43]. Although renowned in healthcare, a consistent number of studies leveraged this methodology for information system research [44]–[46]. Delphi is a multi-round survey procedure aggregating expert opinion, which leads to a consensus. Recent research on the Delphi method also enumerates modifications and adaptations from the original methodology [47]. The first round is usually constituted by an open-ended survey in which experts, sometimes organized as a panel, may provide their opinions or information on a specific content area [48]. A modification allows for some questionnaire items to be predetermined to use the first round to gather additional items, clarify redundancy, or identify issues regarding each statement [49], [50]. Although predetermined content may introduce bias by limiting the topics under analysis, the use of preliminary content generated by a literature review is a widely accepted alternative to the traditional open-ended questionnaire [48], [51]. A further modification also allows for a semi-structured interview to be performed instead of the first open-ended survey [47]. Compared to open-ended surveys, semi-structured interviews allow for a higher level of elaboration from the experts, resulting in a higher amount of qualitative data being available for the study.

## 3.2 **Our Delphi study.**

The modified Delphi study chosen for this research is articulated in three steps. In order to gather the relevant assumptions about blockchain in CE, a review of the literature is done utilizing the Scopus academic database. Only "blockchain" and "circular economy" were used as keywords to keep the research as general as possible. The query performed on Jun. 07 2023, returned 227 downloaded and inspected entries. Given the sample dimension, no time restrictions were used as the earliest article (2017) is relatively recent. By reading abstracts, the authors noted that some articles are sector-specific; however, blockchain experts may not be familiar, for example, with the plastic or energy sector. Therefore, the idea is to keep the analysis at a general level and provide examples of specific sectors where possible. Articles were read carefully, and claims over blockchain integrations in CE were transcribed in a separate word file. Once all the claims were transcribed, they were compared and classified to drop duplicates and identify recurring or unique proposals. The author is aware that, in line with previous modified Delphic studies, this specific step involves a certain degree of arbitrariness, as a different research team would have probably extracted a different number of topics formulated slightly differently. However, it's reasonable to hypothesize that the outcome and validity of the study would not be biased by heterogeneity in the formulation of the items and if a specific integration is divided into one or more items. The ten items extracted from the literature review are summarized in Table 2.

Table 2. Extracted items

| Items | Description | Sources |
|---|---|---|
| Improve logistic processes | Thanks to blockchain's enhanced traceability of transactions in supply chains, stakeholders can improve their inventory efficiency, minimize product and material loss, and positively impact waste production and resource usage. | [52]–[56] |
| Facilitate reverse logistics | Blockchain should ensure the traceability of all transactions in a supply chain so that supply chain stakeholders can easily monitor the reverse logistics process to reuse and refurbish wastes and to return components and unsold products. | [14], [15], [20], [57], [58] |
| Improve supplier selection | Data Transparency on a blockchain can facilitate supplier selection by recording all suppliers' historical performance data. Transparently recorded historical performance can improve supply chain trust, minimize opportunistic behavior, and foster collaboration among stakeholders. | [15], [53], [59], [60] [61] [55] [62] [63] |
| Minimize transaction costs by reducing intermediaries. | By directly connecting buyers and sellers, blockchain technology can lower transaction fees and streamline the exchange of goods, services, or resources. Data transparency can also enable P2P marketplaces as data can be accessible by stakeholders of different supply chains. They may, | [64] [65] [61] [62] [66] [21] |

| | therefore, be able to exchange their waste freely without any middlemen. | |
|---|---|---|
| Incentive circular behavior | Blockchain-based systems can introduce incentives, such as tokens or cryptocurrencies, to reward and encourage circular behavior. | [24] [67] [68] [63] [62] |
| Manage green production. | Blockchain can be used to obtain and store data on green products. For example, blockchain can monitor gas emissions, enabling customers to see whether a product is green or not. Blockchain can audit the quality and safety of the use of chemicals, water, and land (for fashion products, for example). It can also detect working conditions and workers' status by collecting worksite-related data such as light, humidity, temperature, and working hours. | [61] [69] [70] [71][72] |
| Enhance product share and sharing economy. | Blockchain technology may provide networking mechanisms for sharing products among different users. Companies could also share their idle resources to reduce costs and carbon emissions. | [73] [74] [62] [75][76][77] [52] [55] |
| Guarantee authenticity of reused/recycled products. | By leveraging blockchain traceability, consumers can gain confidence about the origin of products, whether they are authentic or not, and also after reuse and recycling. | [78] [61] [52] [70] [21] [59] |
| Prolong product life | Blockchain can be implemented to collect data regarding the lifetime of products/components, use phase, maintenance and repair cycles, and geo-localization. If CE assets are composed of modular components that support forward/backward compatibility, blockchain, and IoT systems can be leveraged to foresee potential equipment breakdown. Repair or recycling policies can then be easily enabled. | [63] [79] [55] [80] [63] |
| Enable Digital Product Passports (DPP) on the blockchain. | Blockchain is sought as an appropriate environment to store digital products passports for being transparent, non-modifiable, and non-erasable | [69] |

In the study's second phase, direct interviews are performed with blockchain experts to evaluate and discuss the topics found in the literature review. Before conducting the interviews, experts received an invitation email with the material concerning the study. The scope of the study was explained, and the content of the questions was disclosed. Experts could then decide whether to participate in the interview, depending on their familiarity and competence with the objects under analysis. Direct interviews with experts, as mentioned above, were semi-structured and lasted an hour on average. In the first phase of the interview, experts were asked, based on a 4-point Likert scale (1= strongly disagree to 4= strongly agree), how much they agreed with the proposed items concerning blockchain integration in CE. In the second phase of the interview, experts were asked to elaborate further on their votes and provide motives for their choices. The author wishes to stress that although the statements are taken and inspired by academic papers, the opinion of the experts on them is not to be in any case intended as an evaluation of the quality of the paper itself. The final claims of the cited studies are not under evaluation in this context.

Being the first phase of the Delphi study substituted with a literature review, the direct interviews become the first phase of convergence [47]. Although suggested by [81], we decided not to encourage participants to propose a new object for the questionnaire. As stressed in [82], this request may pressure participants to alter their views according to the recognized literature. However, further proposals and ideas from experts are discussed and included in the study results.

Building on [83], the comments from the experts will contribute to the discussion phase of the study in which their quotes are elaborated according to the obtained results. Their opinion should enlighten the motive of their selection and further underline if proposals are based on a real technological potential or on false assumptions or misbeliefs. Recommendations from experts are expected to push the research in the right direction.

In the third phase of the study, a report on the first round of responses is sent to experts along with the second survey, which includes the items that didn't reach agreement in the first round. As in [84], the additions are represented by an introductory explanation of the classification and quantitative and qualitative feedback from the first survey for each question. When all experts have responded to the second survey, a final report is elaborated and shared with participants. The author wishes to stress that the report only shows aggregate data, and neither the response of the first or the second survey nor the comments can be traced to the respondent so as to avoid direct influence between participants. Although the list of experts is transparent, in compliance with the Delphi methodology, their response is kept anonymous during and after the study. An outline of the Delphi study is provided in figure 2.

Figure 2. Overview of the Delphi study.

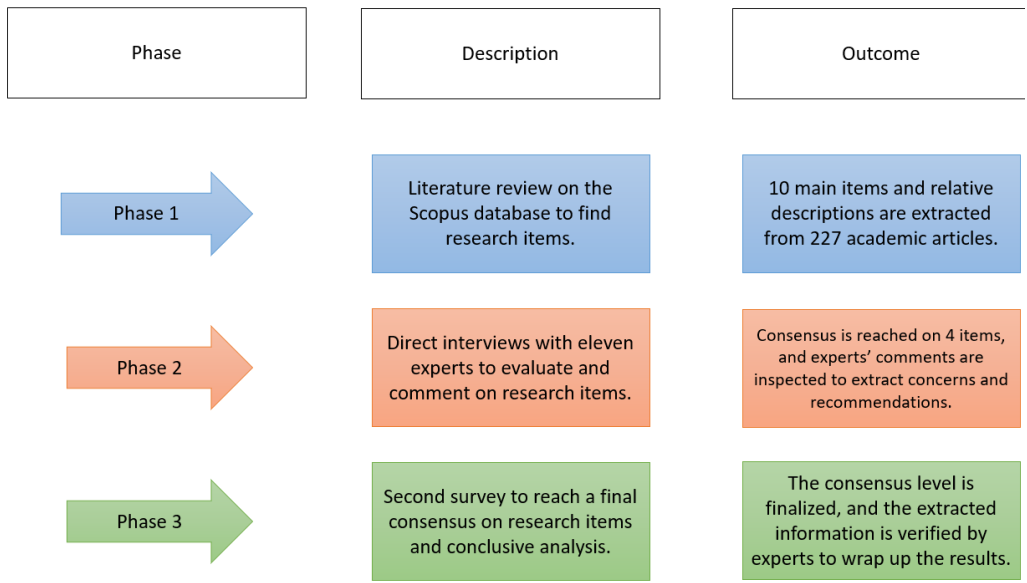

## 3.2 Respondents profile

The selection of experts is critical for the successful outcome of the study. The idea was to differentiate the present research from previous studies in the field by including people who are renowned experts in the blockchain field but are not directly working on circular economy blockchain integrations. This minimizes and hopefully eliminates any bias due to personal conflict of interest. Although this choice may raise concerns about the experts' familiarity with blockchain applications in the CE field, their deep expertise in the blockchain field should provide the necessary guarantees on their awareness of the feasibility of the inspected applications/integrations. Due to that reason and to meet the methodology requirements, the author involved experts with highly heterogeneous backgrounds. No geographical, academic, or professional background restrictions were applied in the expert composition. However, a minimum experience of 5 years in the field is seen as necessary, with a preference of 10 and above. Considering the claims in [4], [32], [85], [86], misconceptions and misbeliefs of blockchain integrations are mainly due to the neglection of oracles and real-world interaction; therefore, the majority of experts interviewed have proven expertise in the oracle field. As indicated by [47], the literature does not show an optimal size of the expert group to take part in a Delphi study. Composed exclusively of highly qualified experts, in line with Dalkey et al. [87], the sample comprises eleven participants to maximize its reliability. The complete list of participants is provided in Table 3.

Table 3. List of participants in the Delphi study.

(The participants' list is made available in the journal version)

### 3.3 Consensus

Concerning the consensus type, as suggested in [82], given the fact that the matter is not of vital importance, such as for healthcare, the consensus may be moderately below 100%. Many studies suggest a level of consensus between 51% and 80% depending on the subject matter, but a percentage above 70% appears to be the most preferred consensus level [81], [88], [89]. Therefore, according to our sample, a desired consensus level for this study would correspond to a minimum of 8 out of 11 respondents agreeing either on a negative (1-2) or positive opinion (3-4) over a specific item. However, a percentage above 51% will be considered already significant (weak consensus). Although respondents have the chance to be neutral (i.e., don't have an opinion), this will not lower the threshold of significance as it will weaken the robustness of the related claims compared to the other items. It has to be stated that to reduce the complexity of the study and reduce the workload of involved experts, the authors chose in advance to limit the number of rounds to two, therefore, not pursuing a consensus for all the items.

### 4. Findings from the second and third phases.

The second phase of the study (first round survey) started on Sept. 21 2023, and finished on Nov. 02 with the interview with the last expert. After the interview round, 4 of 10 topics reached complete consensus (>70%) and were excluded from the next round. Two other items surpassed 51% of the consensus and were still included in the third phase to confirm or improve the consensus level. The third part of the study (second-round survey) was started on November 07, 2023, and finalized by Dec. 09, 2023. Table 4 provides a detailed outcome of the second round. No new items reached consensus during the second round as vote changes were minimal; however, one item passed from a weak consensus to a strong consensus. Finally, six items out of ten reached at least a weak consensus (>51%). From the results, it emerges that forward and reverse logistics didn't reach a consensus, probably due to an abstention rate of almost 30%, although forward logistics is perceived more negatively (1.6 with 45.45%). Supplier selection, instead, reached a strong positive consensus already on the first round (3.375 with 72.72%), while the reduction in transaction costs, although also seen positively, couldn't reach any consensus (3.4 with 45.45%).

Incentivizing circular economy behavior is the item that reached consensus with the highest evaluation (3.75 with 72.72%), while prolonging product life reached the highest consensus and the lowest evaluation (1.25 with 81.81%). Consensus was also reached for supporting green production and for digital product passports with negative opinions, 1.25 and 1.75, respectively. Supporting the sharing economy reached nearly 40% of abstention signs, which is something hard to predict based on the actual state of the art, and authentication support obtained negative opinions but with just a weak consensus (1.83 with 54,54%). Experts' comments were vital to understand the motive of these evaluations, of which an overview is provided in the following paragraphs.

Table 4. Outcome of the second-round survey.

| Item | Mean 1-2 (%) | Mean 3-4 (%) | N° Abst. (%) |
|---|---|---|---|
| Improve logistic processes | 1.6 (45.45%) | 3 (27.27%) | 3 (27.27%) |
| Facilitate reverse logistics | 1.5 (36.36%) | 3.25 (36.36%) | 3 (27.27%) |
| Improve supplier selection | 1.5 (18.18%) | 3.375 (72.72%)** | 1 (9.09%) |
| Minimize transaction costs by reducing intermediaries. | 2 (36.36%) | 3.4 (45.45%) | 2 (18.18%) |
| Incentive circular behavior | 2 (9.09%) | 3.75 (72.72%)** | 2 (18.18%) |
| Manage green production. | 1.25 (72.72%)** | 3.5 (18.18%) | 1 (9.09%) |
| Enhance product share and sharing economy. | 1 (36.36%) | 3.3 (27.27%) | 4 (36.36%) |
| Guarantee authenticity of reused/recycled products. | 1.83 (54.54%)* | 3.3 (27.27%) | 2 (18.18%) |
| Prolong product life | 1.1 (81.81%)** | / | 2 (18.18%) |
| Enable Digital Product Passports (DPP) on the blockchain. | 1.75 (72.72%)** | 3.5 (18.18) | 1 (9.09%) |

*=>51% weak consensus, **=>70% strong consensus.

## 5. Experts' comments, concerns, and recommendations.

The present section provides an overview of experts' comments divided by items. Comments on all the items are included, whether they reached a consensus or not. The reason is to provide a thorough explanation of the reasons why a specific integration is well perceived or not. Also, where agreement is not reached, what motives create further debate. Items follow the same order as Table 2 for better comprehension. Tables 5 to 14 list experts' concerns and recommendations over specific research items.

### 5.1 Improve logistic processes.

Experts' opinion on the use of blockchain for logistic processes is heterogeneous. The few experts that have positive views on the supply chain consider the case of global trade, where there are many actors and massive paperwork with shipping across multiple countries. A blockchain can help harmonize the accounting system by creating a third entry on which all the data and signatures are transparently and immutably stored. However, Many concerns arise when using blockchain to improve traceability and inventory efficiency to minimize material loss and waste production. Improved traceability is not seen as related to blockchain but to oracles and the so-called "stapling problem." The stapling problem concerns the inability to create a stable and trusted link between a physical product and the blockchain [90]. When a shipment arrives, for example, it is impossible to ensure that it exactly corresponds to the one registered on-chain. Furthermore, blockchains are recommended in an environment where parties do not trust each other, while as an expert comments: "*Most of the time when you do business, there is already some trust involved…..the supply chain is still phone calls and handshakes, even faxes. This blockchain optimization is superfluous*". Agreeing with that opinion, another expert sees it as "overengineering" in the sense that in order to achieve full transparency, it is sufficient to publish the data. There is little agreement that the supply chain works poorly for a lack of trust, but this is more due to other factors such as lack of digitalization. Inefficiency in the inventory is mainly due to matters internal to the organization for which an ordinary database is perceived as sufficient. As an expert commented: "*Order or inventory tracking is not a problem as incorrect inventory entry…..human mistake, thefts or accidents can cause products and material loss, blockchain cannot prevent those*". Focusing more on the implementation aspects, another expert commented, "*There are bigger challenges on the implementation rather than usefulness*." He explains that most of the blockchains are not suitable for those applications yet. The reasons are a lack of data interoperability, the difficulty of having everyone make use of one source of truth, and the difficulty of having privacy-preserving smart contracts. Given the setup costs of the blockchain, it would make sense to implement it in the supply chain only if it is already used in the company for something else, but hence, the benefit is perceived as minimal.

Table 5. Experts concerns and recommendations over research items. item #1 blockchain integration in logistic processes.

| Concerns |
| --- |
| There are bigger challenges in the implementation rather than usefulness. |
| Most of the blockchains are not suitable for logistic processes. |
| For real-world assets, blockchain doesn't equal ownership. |
| The same result is achievable by just publishing data. |
| It's still not possible to attach a physical product to the blockchain (Stapling problem) |
| Blockchain is recommended in the absence of trust, while there is always some sort of trust involved in businesses. |
| Products and material loss are not perceived as related to a lack of trust but to a lack of digitalization and human mistakes that blockchain is unlikely to prevent. |
| **Recommendations and conditions** |
| Given the high setup costs, blockchain should only be implemented for internal logistic processes if it is already used for other processes. |
| In the case of global trade, blockchain can simplify the paperwork and signature handling. |
| Blockchain can be suitable to harmonize multiple accounting systems. |

## 5.2 Facilitate reverse logistics.

Although there is no agreement for reverse logistics, we find slightly more positive opinions from experts compared to forward logistic processes. The reason is that another actor is introduced for reverse logistics (the one that sends the shipping back), requiring an additional signature. The coexistence of multiple actors may make the use of blockchain helpful, in theory, although its practical implementation raises a few concerns. Some experts argue that blockchain is unnecessary, as a regular database would suffice. This database can also be distributed among some computers and servers to guarantee a certain amount of decentralization. Slight value is seen as a method to update reverse logistics information in real-time, but again, only in the case that blockchain is already used for some other processes, given the high setup costs. Although one of the experts was very positive about this implementation, he saw it as impractical. The reason is that reverse logistics, in his opinion, is something that companies try to prevent unless there is a market in which waste must be recollected (e.g., batteries). In this case, however, the use of blockchain, in his opinion, is conditional on a decentralized identification service [91].

Table 6. Experts concerns and recommendations over blockchain integration for reverse logistics.

| Concerns |
| --- |
| An ordinary database would suffice. Decentralization is guaranteed by hosting copies by multiple actors. |
| Reverse logistics is not something that companies wish to improve but to prevent. |
| **Conditions and recommendations** |
| Blockchain can be helpful to obtain real-time data on returns. |
| It is recommended when it's already used for other processes. |
| Its integration is conditional on decentralized identity. |
| Recommended for markets where reverse logistics is mandatory (i.e., batteries, medicines) |

### 5.3 Improve supplier selection.

Broad consensus and positive expectations emerge from the chance of applying blockchain for supplier selection. The underlying idea is that, when dealing with a company's reputation, it is preferable to have it stored on a decentralized database. It can create transparency, positively affecting the behavior of suppliers, leading to better collaboration and less fraud. One comment, "*It's a marketing thing. If other entities know you are more reliable, they are more likely to get it from you*". Unlike logistic processes, in this case, there is the need to share the data with others so that blockchain can create value.

Furthermore, since suppliers compete, there is the chance that historical data is tampered with, and blockchain makes sure that data is not manipulated. Other experts agree that decentralized reputation systems will be precious in the future; therefore, it's imperative to prevent centralized entities from being in control of the data. Given the presence of trust, a centralized database is not seen as an appropriate choice, although there is no proof that blockchain is the best alternative. An expert argues that a timestamp with a hash of every data checkpoint could also work, but blockchain natively implements those features. According to some, even a private blockchain can be valuable for this purpose since although some nodes have the power to alter entries, there will always be the "trail" of reordering/reorganization. Further commenting on private blockchains, another expert conjectures that although they are probably more suitable, it is harder to forecast their implementation as: "*is difficult to find people skilled on that, everything has to be built from scratch, it's hard to find tools*." Also, concerns arise for interoperability with current systems.

We remember that due to the Oracle problem, ensuring the truthfulness of on-chain data is challenging, while collusion is possible since multiple parties are involved [32]. However, some experts agree that there is little chance of failure from this point of view. First, as stated before, there will always be the trail of someone who inputs data on the chain, and when recognized as an untrusted entity, all his uploads will be marked as untrusted. Second, if a reputational score is involved and the system is used for a long time, people have little incentive to deviate from honest behavior. However, a more critical opinion states that the signatures of involved parties should accompany the data concerning supplier reviews as proof of the occurred transaction: "*You want to have more than just what the buyer/seller said*".

The only expert against this proposal sees its long-term implementation pessimistically. The underlying idea is that although supplier data is immutable, transparent, and non-erasable, nothing will prevent a user with an honestly earned good reputation from selling the reputation score to another entity, as happens today in web2. Apparently, blockchain-based systems cannot prevent this type of drawback.

Table 7. Experts' concerns and recommendations over blockchain integration to improve supplier selection.

| Concerns |
|---|
| Reputation can be sold, as it happens in web2. Blockchain cannot prevent this type of behavior. |
| A private blockchain is more suitable, but we lack the infrastructure, skills, tools, and interoperability with legacy systems. |
| **Conditions and recommendations** |
| If a blockchain cannot be implemented, a timestamp with a hash at data checkpoints should provide a similar outcome. |
| Above the suppliers' review, all the data and signatures should be stored to certify the genuineness of the transaction that occurred. |

### 5.4 Minimize transaction costs by reducing intermediaries.

A consensus was not reached on this topic, although opinions are not very heterogeneous. Some who had an unfavorable opinion questioned the idea that blockchain reduces transaction fees. One of the experts stated, "*We should just look at the evidence. Has blockchain made anything cheaper?......Blockchain is more expensive than running a database, so the costs will more likely increase*". Others who agree more on the transaction fee reduction specify that this idea only pertains to the area of finance and the possibility of reducing financial intermediaries. It is explained that blockchain with smart contracts can efficiently work as an IPO and substitute lawyers, financial institutions, and centralized markets. Much skepticism and negative opinions emerge when proposing blockchain to intermediate other types of markets. As an expert commented, "*We should be concerned with markets where there is no need for a central intermediary. The electricity market, for example, cannot benefit from blockchain since you will always need a centralized authority to measure electricity consumption and production*". Another expert on this aspect specifies that the problem lies when there is a need for actors/intermediaries to perform operations in the so-called "mid-space" (real-world). In that case, the blockchain being digital cannot operate directly in the real world and cannot guarantee their execution. He further comments that: "*In theory, you can have the perfect blockchain that can make the perfect streamline of uncensored transaction between purchaser and supplier and then you can have unexpected mid actors that preclude buyer to have access to their resources. And blockchain cannot help there at all.*" A more positive opinion from an expert sees the possibility of opening disputes on blockchain for transactions concerning the mid-space that are not executed correctly. For example, if multiple entities control energy meters, an entry on the blockchain can be disputed if collected values are heterogeneous.

Overall however, the shared vision is that there is no reduction of fees or increase in efficiency for transactions that happen in mid-space with the introduction of blockchain. However, another aspect that is underlined by one of the experts concerns the switching costs for operating on the blockchain. Even if operating on the blockchain would be cheaper, there are many complex things to set and learn before using blockchain. Providing the example of a recent experiment, he stated: "*It reminds me of a project named open-bazaar, that was shut down since nobody really used it even if it worked perfectly…..it was not about the saving in transaction cost but the mental cost of using it, running the software, manage the wallet, be the p in p2p*". Therefore, the idea is that the switching costs for learning the new technology may outnumber the exiguous savings in transaction costs.

Table 8. Experts concerns and recommendations over blockchain integration to minimize transaction costs by reducing intermediaries.

| Concerns |
|---|
| There will probably not be a reduction in fees. Maybe an increase. |
| Operations in mid-space cannot be disintermediated with blockchain. |
| The switching costs of learning the new technology may surpass the exiguous reduction in transaction costs. |
| **Conditions and Recommendations** |
| Prioritize the use of blockchain for transactions and operations that can be entirely digitized (i.e., do not need an intermediary that operates in the real-world) |
| Leverage dispute resolution mechanisms, when disintermediating real-world operations. |

**5.5 Incentive circular behavior.**

Except for a negative and two neutral opinions, the expert's sample has high expectations for blockchain to incentivize circular behavior, although many conditions and recommendations are provided, some of which contradict each other. The basic idea is that if a token is issued when a green action is performed, such as recycling a bottle, then the token can be utilized to buy something else, such as a book. Then, the recycling and token issuance can be performed multiple times with different goods, creating and incentivizing a circular economy. As an expert commented, "*In a competitive fashion, people can use tokens to draw and pull resources from other sectors of the economy, as a sort of vampire attack….as it pulls resources from the polluting sectors and puts them into the green sectors*". A vampire attack is a widespread practice in the crypto space and is performed by drawing away customers from a platform offering the same service at a lower cost or higher reward. Although generally considered malpractice, it would be beneficial if used to reduce pollution or waste. The same expert suggests that it would be simpler to organize circular economies through micro-communities issuing their own tokens to reward people for performing a specific job "*in exchange for a promise that may or may not exist.*" The uncertainty is given since the micro-community should be in charge of giving value to the issued token; therefore, a certain amount of trust in the community is necessary. The expert proposes the micro-community because he sees the management of token issuance and resource supervision as extremely unfeasible at a general level. Other experts generally provided similar thoughts on this integration but with the following clarifications. First, an authority, whether centralized or in the form of a DAO, is needed to supervise the operations. The blockchain and smart contracts alone are unable to monitor all the processes. Blockchain as a technology is not strictly necessary to enable the issuance of tokens and recycling processes, but the idea is that it will speed up procedures and provide more transparency. Although agreeing on the general concept, another expert sees it as hard to achieve. The system can work with the right incentive/Tokenomics, but elaborating on the incentive is not easy. He declares, "*If one can design an incentive compatible for all participants that allows these actors to participate in such a circular economy, I would agree more, but I don't know if someone could create that. It is technically feasible but economically difficult.*" Little discrepancies emerge concerning the use and the type of tokens suggested for this application. An expert claims that using BTC or ETH to pay rewards will be unsustainable as it will cause participants to cash out from the system constantly. Likewise, other experts say that a well-designed community token would incentivize users to keep the value in the system, delaying cash-out processes. That way, the system will benefit from closed affiliate programs such as Air Miles and open markets if tokens are tradable on DEXes. Another expert, however, has an opposite vision. He sees, in fact, the coexistence of different tokens for every community or every action as unfeasible since it will create different revenue streams, eventually increasing transaction costs. As he declared: "*I see a good incentive in paying people with crypto that should be the most fungible such as BTC, ETH or DAI*". Another thought concerns the amount of reward per action. An expert, in fact, is skeptical of this opportunity not for the technology itself but for the possible reward amount. In his words: "*They give me some points for recycling, which makes my shopping a little bit cheaper.....I suppose it could be an NFT or a token, but I don't see any benefit... it's such a small incentive!*". This expert declared to be very familiar with these programs as there are many in his hometown, but as he said, they didn't have much traction as the monetary incentive is minimal. So, his idea is that having a non-zero incentive is insufficient, and higher subsidies are needed to enable green actions. This view, however, contrasts with the opinion of the expert who had negative thoughts on this integration. In his view, the higher the incentive, the higher the chance for these ecosystems to be breached. He believes that supervising these operations is unfeasible due to the lack of digitalization. While shipping is digitized and traceable, old clothes, empty cans, or compost are not digitized and, therefore, hard to trace. Many concerns emerge from his words: "*How can I prove that I recycled or reused them? How can I prove that I am recycling compost? Should I weigh it every time? Should we have a smart garbage collector? It would just require such a big change!*" Sharing this skepticism is another expert who proposes a measure to address possible abuses. The incentive to cheat comes from some sort of honey pot where the liquidity for reward is stored and may incentivize fraudsters to manipulate collectors to drain the faucet. The idea of this expert is to embed (where possible) the reward for recycling directly on the product package. Instead of a central authority that monitors the recycling process and rewards for performing some actions, tokens can be locked in the product directly by the producers and then unlocked when they are returned. Therefore, the manipulation possible due to the stapling problem can affect just one reward, not the whole system. The same expert also specifies that manipulations and inefficiencies are not just due to the lack of digitalization. He made an example of "Blockchain Hygiene," in which developers obtained gas refunds if they freed up space on the Ethereum blockchain. It was like an incentive to tidy up the workspace after work. Although useful, it was removed due to abuse. A last incentive that emerges concerns the possibility of getting recognition for donations on circular projects and campaigns. Donors can be rewarded with badges that are not meant to be spent but to identify the source of the donation. With a badge, donors can prove their donation without necessarily breaking their anonymity, enabling further incentive mechanisms.

Table 9. Experts' concerns and recommendations over blockchain integration to incentive circular behavior.

| Concerns |
| --- |
| The incentive is hypothetical since an authority will always be needed to make sure a certain action is rewarded. |
| Token issuance and the economics of the incentive system are seen as hard to design |
| Blockchain and Smart contracts cannot enable these mechanisms alone. |
| There is uncertainty in the token type to use, whether a community one (XYZ) or a global one (BTC, ETH) |
| Non-zero monetary incentives are insufficient to boost circular behavior. |
| The higher the incentive, the higher the chance for the system to be manipulated. |
| Recycling is complicated to incentivize with blockchain due to the lack of digitization. |
| **Conditions and Recommendations** |
| Circular economies should be organized in micro-communities so that behavior can be more efficiently monitored. |
| An authority, whether centralized or in the form of a DAO is required to supervise these operations. |
| A community token may prevent the value from being cashed out from the system, while a global token can allow more fungibility of value for different actions performed. |
| Products should have embedded the amount of money rewarded for their recycling. |
| Blockchain can be leveraged to enable refunds for unused resources, incentivizing efficient resource usage. |
| Donations can be incentivized by issuing tokens through a blockchain that identifies the source of the donation. |

### 5.6 Manage green production.

Comments and opinions from experts on the ability of blockchain to manage and monitor green products and green production are generally negative. Only two experts have a favorable opinion, but both assume that the IoTs work correctly and are not in control of the producing company. Furthermore, one assumes that an external Oracle provider ensures untampered data transmission. From these comments, it is firstly arguable that blockchain cannot operate in this sector without relying on some IoT systems that perform the data gathering. Due to that reason, other experts are highly skeptical. Guaranteeing impartiality and preventing the malfunction of IoTs is seen as something complex, if not impossible, to achieve. As an expert commented: "*With blockchain, you are just using a different database for accounting, not monitoring, these things. There is always an authority that decides how to collect this data and where to put it*". Further, on the trustlessness of these systems, another goes: "*A system based on those sensors should not be trusted, especially when there is a high-value transaction involved, as screwing up these sensors is quite simple.*" Practically is also highly unrealistic to imagine the chance of placing independent sensors. As commented by an expert: "*We should give everyone the possibility to crawl to the chimneys and put the data on the blockchain, which is just not feasible.*"

The way the integration is proposed is then clearly based on some misconception about blockchain potential to independently monitor external state, which is something that pertains to some other technologies that do not bear characteristics such as decentralization, trustlessness, and transparency. More compliant with blockchain potential, although unlikely to be realized, is another idea from some experts. The idea concerns the use of blockchain as a decentralized database, not belonging to any authority, on which anyone (companies, employees, private citizens, and IoT) can independently and freely upload their personally collected data about products and production. Based on the wisdom of the crowd, similar to prediction markets, it may generate more reliable information in real-time. As suggested by an expert, "*it may serve as a sort of diary for workers that remains immutable..... better if both parties are aware of the data uploaded so in case of disputes it's not possible to deny what is declared before....whether is true or not.*" The idea is that blockchain can be a sort of decentralized storage of information concerning working conditions, whose data is agreed on by both parties so that both are protected in case of a dispute. However, preventing collusion or manipulation by a central authority for data concerning specific products or specific productions would still be difficult. Lastly, an expert comments on the fact that instead of using blockchain to monitor green production, the technology can be more efficiently leveraged to tokenize the proof of green production or emission reduction. Some exciting experiments are, in fact, being conducted on the tokenization of carbon credits [92]. The tokenization of carbon credits makes them fungible, allowing their widespread negotiation with higher levels of volumes (including defi platforms), enabling an unpredictable large number of use cases. It may enable, for example, a gateway for liquid carbon credits that companies can burn without any entry barriers. Tokenized carbon credits can also be leveraged to issue wrapped Bitcoins whose issuance can be proven green. Hedge funds that are bound to green investments can then invest in those green Bitcoins, being their clean origin provable on-chain.

Table 10. Experts' concerns and recommendations over blockchain integration to manage green production.

| Concerns |
|---|
| Blockchain cannot monitor external state. |
| IoT data is easy to manipulate. |
| It's unfeasible to extract data such as emissions or pollution in an independent way. |
| **Conditions and Recommendations** |
| IoT systems should be trusted, reliable, and not under the control of a central authority. |
| Blockchain can be used as a database to store real-time data coming from independent data sources. |
| Instead of using blockchain to monitor green production, it can be used to tokenize proof of emission reduction (e.g., carbon credits). |

### 5.7 Enhance product share and sharing economy.

The integration of blockchain for the sharing economy received a balanced distribution of votes among positive, negative, and neutral. It emerged that views are not just heterogeneous, but possible integration is hard to predict. While a few have no opinions, others are just skeptical about using blockchain for the sharing economy, if not for using cryptocurrencies. As an expert commented: "*Apart from the chance of using cryptocurrencies, I don't see how blockchain can improve the sharing economy, especially if we talk about physical products.*" Slightly more enthusiasm is perceived for the sharing of digital assets in which blockchain can help, but another expert commented: "*It will happen once we have the scalability to do it.*" Therefore, it emerges that the technology is not there yet. Concerning the three experts that have positive opinions, their argumentation was the following. The first hypothesized the use of blockchain as a common tech standard that allowed a reduction of the registration and login costs. In his idea, if many different warehouses belonging to different companies have their resources registered on a shared blockchain, they could redistribute them more efficiently. The second suggested the use of blockchain to enable insurance services in case a malfunction is experienced with the shared product. Lastly, one of the experts hypothesizes the use of blockchain to create a decentralized exchange of shared products where users can bid for a specific service. A blockchain is not strictly required to enable this service but will grant openness to it. Furthermore, the blockchain would allow the tokenization and sharing of more significant projects, such as recycling plants where massive capital is needed, and few private entities may be willing to invest in that. As green investments are sometimes not tempting for multinationals or not affordable for the public sector, private individuals can share, thanks to the blockchain, the cost of these investments in their local area.

Table 11. Experts' concerns and recommendations over blockchain integration to enhance product share and sharing economy.

| Concerns |
|---|
| We still don't have the necessary scalability. |
| Blockchain is not strictly needed. |
| **Recommendations and Conditions** |
| Digital products can be more easily shared. |
| It can be used as a common standard for sharing resources (i.e., a decentralized warehouse database) |
| Automated insurance for shared products can be enabled. |
| A sharing economy DEX is feasible with blockchain. |
| It can facilitate green investments by sharing the cost. |

### 5.8 Guarantee authenticity of reused/recycled products.

Opinions of experts on this aspect are heterogeneous, although the majority are not favorable. The main limitation is sought again in the stapling problem, making it impossible to physically attach a product to the blockchain. As an expert commented: "*Anyone can remove that tag and put it to something else…. these systems work with the package, not with the content, so it is hard to imagine a working system for products that get manipulated and transformed during the supply chain*". Also, one of the three experts who had a favorable opinion on authentication is instead dubious about the recycling and reuse part due to possible manipulation. We can argue then that the chance of tracking products after the recycling and the reuse phase with the blockchain is minimal. Concerning the registration of traceability information, the opinion is also quite negative and mainly reflects one of the logistic processes. The general idea is that there is no need for blockchain. Two positive thoughts from experts consider the hypothesis instead of a company that went bankrupt with all the servers and databases dismissed. In that case, having data stored on the blockchain would allow the identification of products throughout their life cycle, regardless of the company's faith. This would be an advantage, especially for collector's items, till the blockchain tag is not manipulated.

Furthermore, an expert considers this application positively not for the intrinsic value of the technology but for the extra effort required by counterfeiters to manipulate the products. He further specified that an intermediary (certification body) is needed to make sure that the data is accurate. However, although those may get bribed, there will always be a track of their action so they can be marked as untrusted. Agreeing on that point, another expert sees these methods as viable only for high-value products for which additional security measures should not affect the final price. For products of low value, the problem is no more the manipulation but the very usability of these tags since end users probably won't even care about them.

Although negative on this integration, an interesting point of view considers the case of products whose parts are made by different suppliers. In this case, the products receive different stamps from different companies, and the likelihood of replication is scarce, although the manipulation or substitution of components is still possible. The concept of substitution is interesting, as the expert comments: "*The swap is not completely free because if you swap a product, you need to have a similar one. You can just swap products; you cannot invent them*". The underlying idea is that, although the tag can be manipulated, a blockchain entry should correspond to a product. Therefore, if, on the one hand, it is possible to substitute a genuine product with a fake one, it is not possible to create counterfeit copies out of thin air. This means that if registered on the blockchain, the total amount of products cannot increase, making it more evident when counterfeit products are added to the market. Further building on this concept, another expert considers the case of products whose copies are hard to distinguish from the original. Artificial diamonds or some designer product clones are extremely difficult to differentiate from the original, even for jewelers. Therefore, the idea is that whoever product is accompanied by the certificate of authenticity becomes the original, so that is the certificate that gives value to the product. That is something seen as easily manageable with blockchain thanks to the NFTs technology.

Table 12. Experts concerns and recommendations over blockchain integration to guarantee authenticity of reused/recycled products.

| Concerns |
| --- |
| It's hard to link a physical product to the blockchain (Stapling Problem). |
| An intermediary (certification body) is needed to make sure that the data is true. The authority can get corrupted and become unreliable. |
| It's impossible to prevent manipulation in the recycle/resale process. |
| The system works with the package, not with the product itself. |
| Considered of low utility for products of exiguous value. |
| **Conditions and Recommendations** |
| For long-term purposes, traceability data can be available forever on a chain. |
| The product should be properly branded, and the tag should not be cloned. |
| More viable for products of high value, given the costs of tags and blockchain fees. |
| Valid, for modular products whose parts are made by different producers |
| For products that are difficult to distinguish from their copies. The authenticity certificate gives value to the underlying product. |

### 5.9 Prolong product life.

Except for a couple of experts who preferred to stay neutral, a broad consensus is reached on the opinion that blockchain unlikely constitutes a compelling use case for prolonging product life. First, it emerges that probably the proposal comes from some misconception about the ability of blockchain to gather data. Blockchain arguably cannot help in performing this task. As explained by experts, data collection is performed by IoT devices that are not yet fully reliable for these types of operations. An expert argues that "*the problem is that IoT devices are not trustless yet. The closest we have are TEES, but if you have physical access to them, they are not that secure*". Above all, the very need for blockchain is questioned in this case. The prediction of component failure is more associated with AI and statistical models that can be more reliable if a high quantity of data is collected by multiple entities and shared in an open database. If a considerable amount of data is collected concerning the failure of the machines, a probability

(conditional on other variables) can be given that the machine will break at a certain point in time. A decentralized marketplace in which information is shared about the previous owner of a machine can also be helpful to predict its life cycle, but the chance for a system like this to take place is seen as minimal.

One of the experts sees theoretically more value in this integration for products such as cars or high-quality smartphones that have many integrated sensors. However, another expert with previous experience in the high-quality smartphone assembly and repair sector questions its practical application. Sensors can hardly initiate transactions to the blockchain, and the cost of handling the fees for storing the considerable amount of data gathered is probably not economically viable. Again, a centralized server can better handle these operations. Arguably, as it is done for ink cartridges, a chip can be placed on the products that communicate with a server to share product information and warn when some product parts need to be changed.

Table 13. Experts' concerns and recommendations over blockchain integration to prolong product life.

| Concerns |
|---|
| Blockchain cannot gather data. It is the IoT that is used for this purpose. |
| IoTs are easy to manipulate if there is direct access to them. |
| Components are owned by the company, and the data they provide cannot be trustless. |
| IoTs cannot initiate transactions as a direct link with the blockchain is hard to establish. |
| Fees would be high due to the considerable amount of data collected by the sensors. |
| Predictions are more associated with AI and statistical models. |
| Conditions and recommendations |
| Appliances with already integrated sensors may be more suitable for this integration (i.e., smartphones) |
| An open blockchain where all product data from many companies can be used as a database for statistical model purposes. |
| Although impracticable, a peer-to-peer marketplace storing all the product data and previous ownership can be helpful in this integration. |

### 5.10 Enable Digital Product Passports (DPP) on the blockchain.

The proposal of having a digital product passport on the blockchain is clearly not supported by experts, as only two had a favorable opinion. However, unlike other proposals, the negative opinion of experts is not due to some technical limitation of blockchain since NFT technology would easily enable DPP to be stored on the blockchain. One of the experts declared: "*Technically, it is possible to store those digital passports on-chain. It is something that is currently done with many documents, but I don't think they'll achieve what they want to with that*". Therefore, mainly conceptual concerns drove negative opinions on that integration. First, there is the issue of credibility of what is written on the digital certificate that is not enhanced by being stored on the blockchain. Then, there is the issue of selecting the most appropriate blockchain. It is perceived as unlikely that an open blockchain such as Bitcoin or Ethereum can be viable for this purpose. The amount of data to be stored is considerable, and the fees are high. So, there is the issue of how to fund the payment of those fees. Then, since it is a European initiative, there is the dilemma of whether to make that information globally accessible or only accessible to some stakeholders. In that case, a private blockchain should be selected, and maintenance costs should be shared somehow. Lastly, there are some more philosophical concerns about whether having those DPP can enhance recycling, as this is more related to human behavior than to lack of information. Apart from the conceptual concerns, some technicalities emerge concerning the very use of blockchain for this. Above the problems with fees and data privacy, there is the idea of immutability, that can't be guaranteed by any blockchain. Many suggest a regular database to reduce costs. The database can then be distributed by hosting multiple copies in any EU state. Another expert suggests having instead a community-driven database such as Wikipedia instead of blockchain, where all the members can contribute. A last interesting consideration from the experts compares the DPP with the concept of a digital passport for humans. Another issue arises since the idea is not just to store product data on the blockchain but to help identify them to facilitate circular economy practices. Digital passports for humans, in fact, are also accompanied by a unique digital signature that can prove the link between the person and the document. A physical, inanimate product cannot provide a signature to prove its identity and connection with the document, limiting much of the potential and use of the underlying technology.

Table 14. Experts' concerns and recommendations over blockchain integration to enable digital product passports on the blockchain.

| Concerns |
|---|
| Blockchain cannot guarantee the authenticity of data provided with digital product passports. |
| It would be unfeasible to leverage open blockchains due to the high costs of fees. |
| It's unclear if transparency or privacy is to be prioritized. |
| If the objective is to guarantee the immutability of certificates, then not all blockchains are non-modifiable and secure. |
| DPPs lose their advantages when used for physical products since, unlike humans, they cannot provide proof of their existence through a signature. |
| **Conditions and recommendations** |
| Multiple parties should verify the DPP. |
| It should be done by a family of products or lots of production to reduce the amount of data to be stored. |
| A community-driven database or a distributed database can also work well. |

## 6. Conclusion

The present paper investigated blockchain potential when integrated with circular economy practices. Delphi technique has been leveraged to evaluate specific topics. Items for the Delphi study were obtained by reviewing academic literature and were evaluated by experts who provided extensive comments to justify their choices. After two rounds of the Delphi study, five items reached broad consensus. Experts see the high potential of blockchain in incentivizing circular economy practice by issuing tokens and facilitating green supplier selection by making their historical performance transparent, immutable, and not manipulable by a central authority. Negative opinions instead emerge on the use of blockchain to monitor green products and production, prolong modular product life, and store digital product passports. Thanks to the expert's comments, further blockchain implementation that can or cannot constitute support for circular practices are also discussed. An overview of these integrations along is provided in table 15.

Table 15. What blockchain can and cannot do for the circular economy according to experts' comments and evaluations.

| Blockchain can |
| --- |
| Incentivize circular behavior through the issuance of tokens or badges. |
| Facilitate the selection of green suppliers, guaranteeing the transparency and immutability of related feedback. |
| Facilitate reverse logistics for material that requires safe disposal or recycling by enabling authentication and transparency of multiple agent's signatures. |
| Facilitate green investments by enabling crowdfunding at reduced transaction costs. |
| Be leveraged to constitute a DAO to manage green initiatives. |
| Tokenize proof of green production or reduction of emissions (i.e., carbon credits). |
| Help harmonize accounting systems (i.e., sustainability accounting) by enabling a triple entry accounting. |
| **Blockchain cannot** |
| Help track and authenticate products after recycling due to manipulation and the stapling problem. |
| Prevents material loss and minimizes waste due to human mistakes. |
| Constitute a convenient platform to store digital product passports due to the high volume of data and transaction fees required. |
| Reduce intermediaries and fees for activities in the mid-space, such as green energy production, water management, etc. |
| Extend product life and facilitate the repairing cycle for the inability to gather and process high data volumes. |
| Monitor green products and productions, being unable to monitor external states efficiently. |

The findings from this paper were useful to better understand the assumptions made in academic literature about real blockchain potential. They can guide manager when developing integration in the circular economy and also investors to have a clearer vision on the most and less promising projects. As advocated by Bockel et al. the study contributes to close the gap between research and practice and provides a robust background to build further studies [1]. This study certainly has some limitations since to avoid biases and conflicts of interests, experts were not directly involved in circular economy initiatives, so they may not be perfectly up to date with the latest advancements in the field. Results are based on experts' opinions and experience and not on a specific use case, therefore, it's plausible that some particular applications may prove to be effective despite negative opinions and vice versa. It is also possible that some integrations have been overlooked in good faith and need further analysis. For example, although blockchain is not usable to reduce intermediaries or costs in green energy production for example, it doesn't imply that it can't be used at all in this sector for other purposes (e.g., tokenize proof of electricity production).

Further studies should verify empirically, leveraging real-world integrations, the expectations from experts to better prove the robustness of this paper results. Also, items that didn't reached consensus need to be further investigated. Lastly it is clearly evident that the list of proposed integrations obtained with the review and analyzed is arguably just a small part of the countless that may be thought in the future. Replication of this study may be necessary to verify the state of the art periodically.